\documentclass[a4paper,10pt]{article}

\def\eq#1{{Eq.~(\ref{#1})}}

\def\refcite{\cite}

\usepackage{latexsym}

\usepackage{graphicx,amssymb}
\usepackage{amsmath}
\newcommand{\cc}{cosmological constant}

\def\mes#1#2{\int_{#2} d^{#1}x\, \sqrt{-g}\, }

\begin{document}


%
%

\title{WHY DOES  GRAVITY IGNORE THE VACUUM ENERGY?}

\author{T. Padmanabhan\\
IUCAA, Pune University Campus,\\
Ganeshkhind, Pune 411007, INDIA\\
email: nabhan@iucaa.ernet.in}
\date{}
\maketitle


\begin{abstract}
The equations of motion for matter fields are invariant under the shift of the matter
lagrangian by a constant. Such  a shift changes the energy momentum
tensor of matter by $T^a_b \to T^a_b +\rho \delta^a_b$. In the conventional
approach, gravity breaks this symmetry and the gravitational field equations
are not invariant under such a shift of the energy momentum tensor. I argue that
until this symmetry is restored, one cannot obtain a satisfactory solution
to the cosmological constant problem. I describe an alternative  perspective to gravity in which
the gravitational field equations  are $[G_{ab}  -\kappa T_{ab}] n^an^b =0$ for all null
vectors $n^a$. This is  obviously invariant under the change $T^a_b \to T^a_b +\rho \delta^a_b$ and restores the symmetry under shifting the matter lagrangian by a constant. 
These equations are equivalent to $G_{ab} = \kappa T_{ab} + Cg_{ab}$
where $C$ is now an integration constant so that the role  of the cosmological constant is very different in this approach. 
The cosmological constant now arises as an integration constant, somewhat like the mass
$M$ in the Schwarzschild metric, the value of which  can
be  chosen depending on the physical context. These equations can be obtained from a variational principle which uses the null surfaces of spacetime as local Rindler horizons and can be given a thermodynamic interpretation. 
This approach turns out to be quite general and can encompass
even the higher order corrections to Einstein's gravity and 
suggests a principle to determine the form of these corrections in a systematic manner. 
  
\end{abstract}


\section{Introduction}

In conventional approach to gravity, one derives the equations of motion
from a lagrangian $L_{\rm tot} = L_{\rm grav}(g) + L_{\rm matt}(g,\phi)$ where
$L_{\rm grav}$ is the gravitational lagrangian dependent on the metric and its derivative
and $L_{\rm matt}$ is the matter lagrangian which depends on both the metric and the 
matter fields, symbolically denoted as $\phi$. This total lagrangian is integrated
over the spacetime volume with the covariant measure $\sqrt{-g} d^4x$ to obtain the 
action. In such an approach, the cosmological constant can be introduced via two different routes
which are conceptually different but operationally the same. 

First, one may decide
to take the gravitational lagrangian to be $L_{\rm grav} =(2\kappa)^{-1}(R-2\Lambda_g)$
where $\Lambda_g$ is a parameter in the  (low energy effective) action  just like
the Newtonian gravitational constant $\kappa$. This is equivalent to assuming that, even in
the absence of matter, flat spacetime is \textit{not} a solution to the field equations.

The second route through  which the cosmological constant can be introduced  is by
shifting  the matter lagrangian by $L_{\rm matt}\to L_{\rm matt} - 2\lambda_m$. The equations
of motion for matter are invariant under such a transformation which implies that --- in the 
absence of gravity --- we cannot determine the value of $\lambda_m$. But such a shift is clearly
equivalent to adding a cosmological constant $2\kappa\lambda_m$ to the
$L_{\rm grav}$. In general, what can be observed through gravitational interaction 
is the combination $\Lambda_{\rm tot} = \Lambda_g
+ 2\kappa\lambda_m$.  

It is clear that there are two distinct aspects to the so called cosmological
constant problem. The first question is why $\Lambda_{\rm tot} $ is very small
when expressed in natural units. Second, since $\Lambda_{\rm tot}$ could have
had two separate contributions from the gravitational and matter sectors, why
does the \textit{sum} remain so fine tuned? This question is particularly relevant because
 phase transitions in the early universe can shift the matter lagrangian
by a large constant; that is, $\lambda_m$ can change during the evolution of the universe.

The transformation $L\to L_{\rm matt} - 2\lambda_m$ shifts in the energy momentum tensor of the form $T^a_b \to T^a_b +\rho \delta^a_b$
with  $\rho=2\lambda_m$ and this is a symmetry
of the matter sector; the matter equations of motion do not care about constant $\rho$.
 \textit{In the conventional approach, gravity breaks this
symmetry.} This is the root cause of the so called cosmological constant problem. As long as
gravitational field equations are of the form $G_{ab} = \kappa T_{ab}$ the theory
cannot be invariant under the shifts of the form $T^a_b \to T^a_b +\rho \delta^a_b$.
Since such shifts are allowed by the matter sector (at least at scales below which
the super symmetry is broken), it is very difficult to imagine a clean solution
to cosmological constant problem within the conventional approach to gravity.
This issue is closely tied to the fact that the action is expressed as
an integral over a bulk spacetime volume with a measure $\sqrt{-g} d^4x$. As 
long as this is the case, the source of gravity will remain $T_{ab}$ and
invariance under the shift $T^a_b \to T^a_b +\rho \delta^a_b$ cannot be 
obtained.

I will outline in this article a new approach to gravity which has promise
to solve this problem. In this approach, the equations of motion are \textit{not} 
$G_{ab} = \kappa T_{ab}$ but $[G_{ab}  -\kappa T_{ab}] n^an^b =0$ for all null
vectors $n^a$. Such an equation is obviously invariant under the shift $T^a_b \to T^a_b +\rho \delta^a_b$ and restores the symmetry under shifting the matter lagrangian by a constant. 
These equations are equivalent to $G_{ab} = \kappa T_{ab} + Cg_{ab}$
where $C$ is now an integration constant. Thus while this approach
leads to  same equations as in 
of Einstein's gravity with cosmological constant, the nature of the constant is very different from 
the usual approach. It is now an integration constant, the value of which  can
be  chosen depending on the physical context --- somewhat like the choice
of $M$ in the Schwarzschild metric.

It is obvious that one cannot obtain the equations $[G_{ab}  -\kappa T_{ab}] n^an^b =0$ from any
conventional action principle. I will show, however, that
 such equations can be obtained from a rather \textit{unconventional} variational principle which uses the null surfaces of the spacetime in an essential manner. In this approach any bulk cosmological constant will become irrelevant. 

The plan of the paper is as follows: I will begin with a broad, rapid, overview
of the current cosmological paradigm in order to present the case for the 
cosmological constant as a choice for the dark energy. I will then describe
briefly the new perspective in the context of general relativity and explain
how it offers new insights into understanding several issues in horizon thermodynamics.
In Sec.\ref{sss:gb}, I will show that this approach is actually quite general and can allow us to determine higher order corrections to Einstein's gravity in a systematic manner. 

\section{The evidence for the cosmological constant}

It is conventional to measure
the energy densities of the various species which drive the expansion of the universe in terms of a \textit{critical energy density} $\rho_c=3H^2_0/8\pi G$  where $H_0=(\dot a/a)_0$
is the rate of expansion of the universe at present.  The variables $\Omega_i=\rho_i/\rho_c$ 
will then give the fractional contribution of different components of the universe ($i$ denoting baryons, dark matter, radiation, etc.) to the  critical density. Observations  suggest that the universe has
 $0.98\lesssim\Omega_{tot}\lesssim1.08$  with radiation (R), baryons (B), dark matter, made of weakly interacting massive particles (DM) and dark energy (DE) contributing  $\Omega_R\simeq 5\times 10^{-5},\Omega_B\simeq 0.04,\Omega_{DM}\simeq 0.26,\Omega_{DE}\simeq 0.7,$ respectively. All known observations\cite{cmbr,baryon,sn,snls}
are consistent with such an --- admittedly weird --- composition for the universe. 

The conventional cosmological paradigm --- which is remarkably successful --- is based on these numbers and can be summarised\cite{adcos} as follows:
The  key idea is that if there existed small fluctuations in the energy density in the early universe, then gravitational instability can amplify them in a well-understood manner  leading to structures like galaxies etc. today. The most popular model for generating these fluctuations is based on the idea that if the very early universe went through an inflationary phase\cite{inflation}, then the quantum fluctuations of the field driving the inflation can lead to energy density fluctuations\cite{genofpert,tplp}. It is possible to construct models of inflation such that these fluctuations are described by a Gaussian random field and are characterized by a power spectrum of the form $P(k)=A k^n$ with $n\simeq 1$. The models cannot predict the value of the amplitude $A$ in an unambiguous manner but it can be determined from CMBR observations. The CMBR observations are consistent with the inflationary model for the generation of perturbations and gives $A\simeq (28.3 h^{-1} Mpc)^4$ and $n\lesssim 1$. (The first results were from COBE \cite{cobeanaly} and
WMAP has re-confirmed them with far greater accuracy).
When the perturbation is small, one can use well defined linear perturbation theory to study its growth. But when $\delta\approx(\delta\rho/\rho)$ is comparable to unity the perturbation theory
breaks down. Since there is more power at small scales, smaller scales go non-linear first and structure forms hierarchically. 
The non linear evolution of the  \textit{dark matter halos} (which is an example of statistical mechanics
 of self gravitating systems; see e.g. ref.\refcite{smofgs}) can be understood by simulations 
 as well as theoretical models based on approximate 
 ansatz\cite{nlapprox} and  nonlinear scaling relations\cite{nsr}.
 The baryons in the halo will cool and undergo collapse
 in a fairly complex manner because of gas dynamical processes. 
 It seems unlikely that the baryonic collapse and galaxy formation can be understood
 by analytic approximations; one needs to do high resolution computer simulations
 to make any progress\cite{baryonsimulations}. The results obtained from all these
 attempts are broadly consistent with observations. 
 
 So, to the zeroth order, the universe is characterized by just seven numbers: $h\approx 0.7$ describing the current rate of expansion; $\Omega_{DE}\simeq 0.7,\Omega_{DM}\simeq 0.26,\Omega_B\simeq 0.04,\Omega_R\simeq 5\times 10^{-5}$ giving the composition of the universe; the amplitude $A\simeq (28.3 h^{-1} Mpc)^4$ and the index $n\simeq 1$ of the initial perturbations. Validating such a cosmological paradigm by different observations is a  remarkable progress by any sensible criterion.
 
 The remaining challenge, of course,  is to make some sense out of these numbers from a more fundamental point of view. Among all these components, the dark energy, which exerts negative pressure, is probably the weirdest.  To understand its rapid acceptance by the community one needs to look at its recent history briefly.  Early analysis of several 
 observations\cite{earlyde}
indicated that this component is unclustered and has negative pressure --- the observation which made me personally sit up and take note being the APM result. This is confirmed dramatically by the supernova observations\cite{sn,snls}. (For a critical look at the current data, see ref. [\refcite{tptirthsn1}]). 
 The key observational feature of dark energy is that --- treated as a fluid with a stress tensor $T^a_b=$ dia     $(\rho, -p, -p,-p)$ 
--- it has an equation of state $p=w\rho$ with $w \lesssim -0.8$ at the present epoch. 
The spatial part  ${\bf g}$  of the geodesic acceleration (which measures the 
  relative acceleration of two geodesics in the spacetime) satisfies an \textit{exact} equation
  in general relativity  with the source being $(\rho + 3p)$.
  As long as $(\rho + 3p) > 0$, gravity remains attractive while $(\rho + 3p) <0$ can
  lead to repulsive gravitational effects. In other words, dark energy with sufficiently negative pressure will
  accelerate the expansion of the universe, once it starts dominating over the normal matter.  This is precisely what is established from the study of high redshift supernova, which can be used to determine the expansion
rate of the universe in the past \cite{sn,snls}. 

The simplest model for  a fluid with negative pressure is the
cosmological constant (for a sample of recent reviews, see ref.\refcite{cc}) with $w=-1,\rho =-p=$ constant.
If the dark energy is indeed a cosmological constant, then it introduces a fundamental length scale in the theory $L_\Lambda\equiv H_\Lambda^{-1}$, related to the constant dark energy density $\rho_{_{\rm DE}}$ by 
$H_\Lambda^2\equiv (8\pi G\rho_{_{\rm DE}}/3)$.
In classical general relativity,
    based on the constants $G, c $ and $L_\Lambda$,  it
  is not possible to construct any dimensionless combination from these constants. But when one introduces the Planck constant, $\hbar$, it is  possible
  to form the dimensionless combination $H^2_\Lambda(G\hbar/c^3) \equiv  (L_P^2/L_\Lambda^2)$.
  Observations then require $(L_P^2/L_\Lambda^2) \lesssim 10^{-123}$.
  As has been mentioned several times in literature, this will require enormous fine tuning. 

Because of this conceptual problem associated with the cosmological constant, people have explored a large variety of alternative possibilities. The most popular among them uses a scalar field $\phi$ with a suitably chosen potential $V(\phi)$ so as to make the vacuum energy (as well as the parameter $w$) vary with time like in quintessence\cite{phiindustry}, kessence\cite{kessence}, tachyonic models\cite{tptachyon,tachyon} etc. 
The hope then is that, one can find a model in which the current value can be explained naturally without any fine tuning. While the scalar field models enjoy considerable popularity 
  it is very doubtful whether they have helped us to understand the nature of the dark energy
  at any deeper level. These
  models, viewed objectively, suffer from several shortcomings:
 
 (a) They completely lack predictive power. It can be explicitly
  demonstrated\cite{tptachyon,ellis} that,  virtually every 
  form of $a(t)$ can be modeled by a suitable ``designer" $V(\phi)$.
 
 (b) These models are  degenerate in another sense.  Even when $w(a)$ is known/specified, it is not possible to proceed further and determine
  the nature of the scalar field Lagrangian. The explicit examples given in the literature show that there
  are {\em at least} two different forms of scalar field Lagrangians, corresponding to
  the quintessence or the tachyonic field, which could lead to
  the same $w(a)$. (See the first paper in ref.\refcite{tptirthsn1} for an explicit example of such a construction.)

 (c) By and large, the potentials  used in the literature have no natural field theoretical justification. All of them are non-renormalisable in the conventional sense and have to be interpreted as a low energy effective potential in an ad hoc manner.
 
 (d) One key difference between \cc\ and scalar field models is that the latter lead to a $w(a)$ which varies with time. If observations have demanded this, or even if observations have ruled out $w=-1$ at the present epoch,
  then one would have been forced to take alternative models seriously. However, all available observations are consistent with \cc\ ($w=-1$) and --- in fact --- the possible variation of $w$ is strongly constrained\cite{jbp}. 
 
 (e) All the scalar field potentials require fine tuning of the parameters in order to be viable. This is obvious in the quintessence models in which adding a constant to the potential is the same as invoking a \cc. So to make the quintessence models work, \textit{we first need to assume the \cc\ is zero.} These models, therefore, merely push the cosmological constant problem to another level, making it somebody else's problem! 
 
The last point makes clear that if we shift $L\to L_{\rm matt} - 2\lambda_m$ in an otherwise successful scalar field model for dark energy, we end up `switching on' the cosmological constant and raising the problems again. It is therefore important to address this issue, which we will the task we address in the rest of the paper.

\section{Gravity's immunity from vacuum energy}
\label{sss:gravimmu}

As we said before the vacuum energy density, regularized at a length scale $L$ will lead to 
a $\rho\approx L^{-4}$. Current observations suggest that something very similar to vacuum energy density --- with $T^a_b=L^{-4}\delta^a_b; L=L_{obs}\approx 0.1$mm --- is producing gravitational effects in the large scale dynamics of the universe. 
We also have reason to believe that our universe went through several phase transitions in the course of its  evolution, each of which shifts the energy momentum tensor of matter by $T^a_b\to T^a_b+L^{-4}\delta^a_b$ where $L$ is the scale characterizing the transition. For example, the GUT and Weak Interaction scales are about $L_{GUT}\approx 10^{-29}$ cm, $L_{SW}\approx 10^{-16}$ cm which are tiny compared to $L_{\Lambda}$. 
Even if we take a more pragmatic approach, the observation of Casimir effect in the lab sets a bound that $L<\mathcal{O}(1)$ nanometer, leading to a $\rho$ which is about $10^{12}$ times the observed value\cite{gaurang}. Given all these, it seems reasonable to assume that gravity is quite successful in ignoring most of the energy density in the vacuum. 

The key issue is that \textit{gravity} is again acting as an odd-man-out. All other interactions of nature are  invariant under shifting $L_{matter}$ by a constant which  shifts the energy by $T_{ab}\to T_{ab}+\Lambda g_{ab}$. But since the matter lagrangian couples to gravity through a $L_{matter}\sqrt{-g}$ term the equations of motion for gravity are \textit{not} invariant under
 shifting the $L_{matter}$ by a constant.  This  problem  will persist in any theory of gravity with equations of motion of the form
$E_{ab}= T_{ab}$ where $E_{ab}$ is some divergence-free, symmetric, tensor made from the metric and derivatives. This, in turn, arises because we use an action functional for gravity with an integral over bulk spacetime volume
with the measure $\sqrt{-g}d^Dx$.

So the only way out of this problem  
is to change field equations\cite{cc1} such that they become invariant under $T_{ab}\to T_{ab}+\Lambda g_{ab}$. This is same as working with the trace-free part of the equations (as originally attempted in the unimodular theories of gravity; see ref.\refcite{unimod})  or, alternatively, with the equations
\begin{equation}
(G_{ab}-8\pi T_{ab})\xi^a\xi^b=0
\label{nulleq}
\end{equation} 
for all \textit{null} vectors $\xi^a$. Either formulation, when combined with Bianchi identity, leads to $G_{ab}=8\pi T_{ab} +C g_{ab}$ with some undetermined integration constant $C$. The $C$  is no longer a coupling constant in the field equations (that is, it does not exist in the lagrangian of the theory) but is part of the solution --- like $M$ in the Schwarzschild metric. One is free to choose it differently in different contexts, depending on physical situation. 
While this does not ``solve" the cosmological constant problem, it changes its nature completely because the theory is now invariant under $T_{ab}\to T_{ab}+\Lambda g_{ab}$.

As long as we take the action for gravity to be an integral over a local Lagrangian density, one cannot obtain  of the equations of motion which are invariant under the shift
$T_{ab}\to T_{ab}+\Lambda g_{ab}$.
 Obviously, the conventional action principle with the gravitational degrees of freedom residing in the bulk cannot give raise to \eq{nulleq}. But if we have only surface degrees of freedom, it seems plausible that the gravity will be unaffected by bulk vacuum energy. [In fact, the 
shift from volume degrees of freedom to area degrees of freedom  can change\cite{cc1,cc2}
  the effective 
  energy density of the vacuum that is coupled to gravity from the gigantic $L_P^{-4}$ to the observed value
 $L_P^{-4} (L_P^2/S)$ with $S\approx H^{-2}$ and can lead to the observed value of the cosmological constant.]

\section{Gravity from the surface degrees of freedom}
\label{sss:holo}  

As I mentioned above one cannot obtain  \eq{nulleq} from a bulk action which is an integral of a local lagrangian $L$ with a measure $\sqrt{-g}d^Dx$. It is, however, well known that
Einstein-Hilbert Lagrangian  $L_{EH}\propto R$ has a formal structure
\begin{equation}
 L_{EH}\sim R\sim (\partial g)^2+\partial^2g \equiv \sqrt{-g}L_{\rm bulk} - L_{\rm sur} 
 \label{one}
\end{equation}
with
\begin{equation}
L_{\rm bulk}=2Q_{ab}^{dc}g^{bi}\Gamma^a_{dk}\Gamma^k_{ic};\qquad
L_{\rm sur}
=2Q_{ak}^{cd}\partial_c\left[\sqrt{-g}g^{bk}\Gamma^a_{bd}\right]
\label{sep1}
\end{equation} 
where $2Q_{ab}^{cd}=(\delta^d_a\delta^c_b-\delta^c_a\delta^d_b)$ is the alternating (`determinant') tensor.
(We have introduced a minus sign in the surface term \eq{one} for future convenience; with this convention, $L_{sur}$ is the lagrangian that should be \textit{added} to $L_{EH}$ to get an action which is quadratic in the first derivatives of metric.)
The surface term obtained by integrating $L_{sur}\propto \partial^2g$ should be ignored (or, more formally, canceled by an extrinsic curvature term; see e.g. [\refcite{gh}]) to obtain 
a well defined variational derivative that will lead to Einstein's equations.
So the (covariant) field equations  essentially arise from the variation of the non-covariant (or foliation dependent) bulk
term $L_{bulk}\propto (\partial g)^2$ --- usually called the $\Gamma^2$ Lagrangian. 

On the other hand, there exists a remarkable relation \cite{tpholo,tpreview} between these two parts of the Lagrangian in the Einstein-Hilbert action: 
\begin{equation}
[(D/2)-1]\sqrt{-g}L_{sur}=\partial_a\left(g_{ik}\frac{\partial \sqrt{-g}L_{bulk}}{\partial(\partial_a g_{ik})}      \right)
\label{surbulkrel}
\end{equation} 
(The result is quoted for $D>4$ for future convenience; of course, for $D=4$, the numerical coefficient in the left hand side becomes unity.) Given this relation, the transition from $L_{bulk}$ to $L_{EH}=L_{bulk}-L_{sur}$ can  be thought of as a transition to momentum representation.  Given any \(L_q(\dot q,q)\), we can always construct a \(L_p(\ddot q,\dot q, q)\) which \textit{depends on the second derivatives} $\ddot q$ but gives the same equation of motion
  by using
\begin{equation}
 L_p=L_q-\frac{d}{dt}\left(q\frac{\partial L_q}{\partial\dot q}\right)
\end{equation} 
   Keeping     {{\(\delta p = 0\)}} at the end points and   varying \(L_p\)
 leads to the the same equations of motion  as  keeping  {{\(\delta q =0\)}} at the end points and
 varying \(L_q\).  
 In quantum theory, the path integral with \( L_p\)  gives the momentum space kernel \(G(p_2,t_2;p_1,t_1)\) just as path integral
 with \( L_q\)  gives the co-ordinate space kernel \(K(q_2,t_2;q_1,t_1)\).
Relation of this kind clearly indicates that both $L_{sur}$ and $L_{bulk}$ contain the same information content.  This strongly suggests that one may be able to obtain the same 
information about the dynamics just from the surface term $L_{sur}$ alone. (For this reason, we shall call this relation \textit{holographic}, where the term is used in a specific sense, as explained in ref. \refcite{ayan}.)
Such a perspective turns out to be remarkably resilient and useful. I will briefly indicate how standard Einstein's theory arises from an  action 
 functional\cite{newper} containing \textit{only} a surface term (delegating the details to the Appendix) and then I will develop a general frame work for the low energy gravitational action functional which has a similar holographic structure. 

\subsection{A new variational principle} 
 
To do this, let us construct a total action for gravity and matter\cite{newper} by adding to $A_{sur}$ the matter action; that is, 
\begin{equation}
A_{tot}=A_{sur}+A_{matter}[\phi_i,g]
\end{equation}
where $A_{matter}[\phi_i,g]$ is the standard matter action  in a spacetime with metric
$g_{ab}$. The $\phi_i$ denotes some generic matter degrees of freedom; varying $\phi_i$
will lead to standard equations of motion for matter in a background metric and these equations will also ensure that the energy momentum tensor
of matter $T^a_b$ satisfies $\nabla_aT^a_b=0$.
Consider now the variation of 
$A_{tot}$ when the metric changes by $g^{ab}\to g^{ab}+\delta g^{ab}$ with $\delta g^{ab}=\nabla^a\xi^b+\nabla^b\xi^a$
where $\xi^a$ is (at present) an unspecified vector field. We keep all other matter variables unchanged.
 Direct calculation shows that (see \eq{appdeltot} of Appendix):
\begin{equation}
16\pi\delta A_{tot}
=2\int_{\partial\mathcal{V}}d^{3}x\sqrt{h}
(R^a_b-8\pi T^a_b)\xi^bn_a
-\int_{\partial\mathcal{V}}d^{3}x\, \sqrt{h}\, (n_cM^c_{\phantom{c}lm}\delta g^{lm})
\label{deltot}
\end{equation}
where $M^c_{\phantom{c}lm}$ is given by \eq{mclmeqn} in the Appendix (the explicit form of which is irrelevant for our purpose) and $n_c$ is the normal to the surface $\partial \mathcal{V}$.
I stress that we are \textit{not} introducing a coordinate shift $x^a\to \bar{x}^a=x^a+\xi^a$ (diffeomorphism) but merely restricting ourselves to a specific type of $\delta g^{ab}$ parametrized by a vector field $\xi^a$.  Under a coordinate transformation $x^a\to \bar{x}^a=x^a+\xi^a$ \textit{all} tensorial quantities change --- not just the metric. In the matter sector, there will be extra variations arising from the changes induced in the matter action due to variations in matter fields $\phi_i$. 
Instead we are merely considering the variation of the total action for a specific type of variation of the metric. The derivation of \eq{appdeltot} in the Appendix makes this clear. 

This variation will lead to sensible equations of motion in a special circumstance which I will now explain.
Around  any event $\mathcal{P}$ in the spacetime  one can introduce a local inertial frame such that the null surface of light rays passing though $\mathcal{P}$ takes the usual $\textbf{X}^2=T^2$ form. One can now transform to a (family of) Rindler frame(s) from this locally inertial frame (in fact with any proper acceleration one likes)
such that the null surface acts as a local horizon for these observers. The
 null surface through at any event on the spacetime can be thought of as a local Rindler horizon for a suitably chosen congruence of observers\cite{probe}. 
We now demand that, whenever the boundary ${\partial\mathcal{V}}$
has a part $\mathcal{H}$ which is a local Rindler horizon, the contribution to the variation of 
 the action from that $\mathcal{H}$ should vanish if
 $\xi^a$ is the Killing vector that generates  $\mathcal{H}$. (There is strong physical motivation for this demand which I have given in previous works, e.g., ref.\refcite{tpholo}; I will not repeat it here.)
 The fact that $\xi^a$ is the Killing vector implies that $\delta g^{ab}=0$ on the Killing horizon $\mathcal{H}$ making the second term in \eq{deltot} vanish. Since the normal to $\mathcal{H}$ is in the direction of 
$\xi^a$, the first term gives
\begin{equation}
(R^a_b-8\pi T^a_b)\xi^b\xi_a=0
\label{firststep}
\end{equation} 
One can do this around every event in spacetime locally  and hence this result should hold everywhere. Using the fact that $\xi^a$ is arbitrary \textit{except for being a null vector}, this requires  $R^a_b-8\pi T^a_b=F(g)\delta^a_b$, where $F$ is an arbitrary function of the metric. 
Writing this as $(G^a_b - 8\pi T^a_b ) = Q(g) \delta^a_b $ with $Q= F - (1/2) R$ and using $\nabla_a G^a_b = 0 , \nabla_a T^a_b =0$ we get $\partial_bQ=\partial_b [ F - (1/2) R] =0$;
 so that $Q$ is an undetermined integration constant, say $\Lambda$, and $F$ must have the form
$F=(1/2)R+\Lambda$. 
The resulting equation is
\begin{equation}
R^a_b-(1/2)R\delta^a_b=8\pi T^a_b+\Lambda\delta^a_b
\label{eom}
\end{equation}
which leads to Einstein's theory \textit{with a cosmological constant appearing as an integration constant} \cite{cc1,newper}. Since this should hold at every null surface through every event, the field equations hold at every event.

This formalism is quite different from conventional action principles --- in which surface terms cannot lead to anything nontrivial --- but ties up neatly with several conceptual issues which I shall briefly address:
 
(a) The surface term of
 gravitational action principle has a thermodynamic interpretation:
$A_{sur}$ is directly related to the (observer dependent horizon) entropy. For example, if we choose a local Rindler frame near the horizon with the Euclidean continuation for the metric:
\begin{equation}
ds^2_E\approx N^2 d\tau^2 +dN^2/\kappa^2+dL_\perp^2
\end{equation}  
then horizon maps to the origin and the region outside the horizon corresponds to $N>0$. 
This fits with our idea that  observers with a horizon should only use regions and variables accessible to them. The surface term can now be computed by integrating over the surface 
$N=\epsilon, 0<\tau<2\pi/\kappa$ and taking the limit  $\epsilon\to 0$. This calculation gives
\begin{equation}
A_{\rm sur}=-\frac{1}{4}\frac{\mathcal{A}_\perp}{G}
\end{equation} 
where $\mathcal{A}_\perp $ is the area in the transverse directions.
Since the surface contribution is due to the existence of an inaccessible region we can identify $(-A_{\rm sur})$ with an entropy. 

(b) It is this entropy we are varying in our variational principle.  Our variational principle has nontrivial content when we choose the boundary ${\partial\mathcal{V}}$ to have a piece made of local Rindler  horizon $\mathcal{H}$ around some event. We then choose $\xi^a$ to be the local Killing vector generating the Rindler horizon and demand that the contribution to the variation of action in \eq{deltot} from $\mathcal{H}$ should vanish. 
 This is a purely local construction on a small patch of ${\partial\mathcal{V}}$, which is a null surface $\mathcal{H}$
(viz. the local Rindler horizon)
using local Killing vector $\xi^a$ which generates it. These exist at all events in spacetime and since ${\partial\mathcal{V}}$ is arbitrary,
this construction leads to the validity of the resulting equations at all events in spacetime. In particular, I stress that: (a) We are not assuming anything about the global structure of the spacetime or the existence of any global Killing vector etc. (b) In the integral over 
${\partial\mathcal{V}}$, there will be contribution from the small patch $\mathcal{H}$ as well as from the rest of the boundary. We only require that the contribution from $\mathcal{H}$ should vanish when it is the local Rindler horizon generated by a local Killing vector $\xi^a$.
It is this local nature which allows us to constrain the integrand without worrying about the contributions from rest of ${\partial\mathcal{V}}$. This role of surface terms in the action functional, when the surface acts as local Killing horizon,  is   supported by the identification of the surface term with
of horizon entropy. Since any generic null surface can act as a horizon for \textit{some} class of observers, this again suggests that the physically relevant gravitational degrees of freedom reside in the surface.  (We will discuss the role of null surfaces in slightly more detail later on.)

(c) The special kind of variation we considered is closely related to the changes induced by a virtual, infinitesimal, 
displacement of the horizon normal to itself. This leads to a  change in the entropy $dS$ due to virtual work in the membrane paradigm of horizons \cite{membrane}. The variation
 of the matter term contributes the $PdV$ and $dE$ terms and the entire variational principle is equivalent to the thermodynamic identity 
$TdS=dE+PdV$ applied to the changes when a horizon undergoes a virtual displacement. 
In the case of spherically symmetric spacetimes with $g_{00}=1/g_{rr}=-f(r)$ this is easy to demonstrate this result explicitly\cite{ss}. The broad, necessarily speculative picture that emerges is clearly the one in which the
 continuum spacetime is like an elastic solid (`Sakharov paradigm'; see e.g. ref. \refcite{sakharov}) with Einstein's equations providing the macroscopic description.

It may be noted that several possible surface actions may have the same variation for certain class of surfaces. 
In the literature one often uses another (Gibbons-Hawking) surface term, which involves the trace of the extrinsic curvature. The difference between these two are explained in the Appendix.
As long as one is led  to the correct equations, classical theory cannot distinguish between different surface terms. 
Our particular choice is dictated by the thermodynamic interpretation mentioned above.

\subsection{Entropy of null surfaces}

 It is possible to arrive at the same result in a different manner which highlights this aspect in a somewhat more conventional way\cite{elastic}. Consider a spacetime with a metric $g_{ab}$ and matter energy momentum tensor $T_{ab}$. Such a spacetime will have several null surfaces, the normals to which will be null vector fields defined on the spacetime. Each of these null surfaces will act as a horizon to a set of local Rindler observers. Based on this consideration, we will associate to any such null vector field $v^a(x)$ an  the entropy functional given by: 
      \begin{eqnarray}
      S&=&\frac{1}{8\pi}\int d^4x\, \sqrt{-g}\, \left[
      2Q^{cd}_{ab}  \nabla_c v^a  \nabla_d v^b + 8\pi T_{ab} v^av^b\right]\nonumber\\
      &=&\frac{1}{8\pi}\int d^4x\, \sqrt{-g}\, \left[(\nabla_a v^b)(\nabla_b v^a) - (\nabla_b v^b)^2
       + 8\pi T_{ab} v^av^b\right]
      \label{freeenergy}
      \end{eqnarray}
      We now demand that this entropy is maximised for \textit{all null vectors} fields $v^a$ in the spacetime. To take into account the null condition we introduce the  lagrange multiplier $F(g) v^av_a$ where $F$ is at present an arbitrary functional of the metric.
      Extremising $S$ with respect to the null vector field $v^a$ will lead to the equation
      \begin{equation}
      (\nabla_a\nabla_b - \nabla_b\nabla_a) v^a = \left(8\pi T_{ab} +F(g) g_{ab}\right) v^a
      \end{equation}
      The left hand side is $R_{ab}v^a$ due to the standard identity for commuting the covariant
      derivatives. Hence an analysis similar to the one given earlier for \eq{firststep} will now lead to \eq{eom} i.e., to Einstein's equations with an arbitrary cosmological constant \cite{othernull}.
       
       Note that we did {\it not} vary  the metric tensor in \eq{freeenergy} to obtain our field equations. In this approach,
$g_{ab}$ and $T_{ab}$ are derived macroscopic quantities and are not fundamental variables. Einstein's equations arise as a consistency condition for \textit{all} null surfaces in the spacetime to have maximum entropy. The action in \eq{freeenergy} is explicitly invariant under $T_{ab}\to T_{ab}+\rho g_{ab}$, which is an idea we started with to facilitate gravity to ignore the cosmological constant.

The expression for the entropy in Eq.(\ref{freeenergy}) 
      reduces to a four-divergence   when Einstein's equations
      are satisfied (``on shell") making $S$  a surface term: 
      \begin{equation}
      S = \frac{1}{8\pi}\int_{\cal V} d^4x\, \sqrt{-g}\, \nabla_i ( v^b \nabla_b v^i - v^i \nabla_b v^b)
      =\frac{1}{8\pi}\int_{\partial{\cal V}} d^3x\, \sqrt{h}\, n_i( v^b \nabla_b v^i - v^i \nabla_b v^b)
\label{onsur}
      \end{equation}
      The entropy of a bulk region $\mathcal{V}$  of spacetime  resides in its boundary
      $\partial \mathcal{V}$ when Einstein's equations are satisfied. (This is the motivation for calling that functional entropy, in the first place). In varying  Eq.(\ref{freeenergy})
      to obtain the field equations, we keep this surface contribution to be a constant. This is  the idea we used earlier to obtain Einstein's equation by directly varying a surface term in action. If the spacetime has microscopic degrees of freedom,
then any bulk region will have an entropy and it has always been a surprise why the entropy scales as
the area rather than volume. Our analysis shows that, the semiclassical limit, when Einstein's equations hold
to the lowest order, {\it the entropy is contributed only by the boundary term.}  It is easy to show that, in this case, we will get an entropy that is
       proportional to the area of any horizon, if the horizon arises as a singular point in the
       null vector field.

\subsection{Symmetries of the surface term} 

Since this approach has brought to center-stage the surface term, it is worth pointing out an important property of this term, viz that it is nonperturbative in the gravitational coupling constant. In a general gauge, the $L_{sur}$ of the Einstein-Hilbert action has the form:
\begin{equation}
\sqrt{-g}L_{sur}=\frac{1}{2l_P}\partial_a[\sqrt{-g}(g^{ab}\Gamma^c_{bc}-g^{bc}\Gamma^a_{bc})]
\label{exactsur}
\end{equation} 
Consider now the linear approximation to Einstein gravity around flat spacetime
("graviton in flat spacetime") obtained by taking $g_{ab}=\eta_{ab}+l_P h_{ab}$ with $l_P^2=8\pi G$. (The dimension of $h_{ab}$ is $1/L$ as expected.) Then the Hilbert action has the structure:
\begin{equation}
 \sqrt{-g}L_{EH}=\frac{1}{2 l_P^2}\sqrt{-g}R\sim\frac{1}{2l_P^2}[(\partial g)^2+\partial^2g]
=\frac{1}{2}(\partial h)^2+\frac{1}{2l_P}\partial^2 h
\end{equation}
We see that the surface term is non-perturbative in $l_P$ in the ``graviton" picture! It follows that
staring from quadratic spin-2 action and iterating to all orders in the coupling constant $l_P$ can \textit{never} lead to a term which is non-analytic in $l_P$. So such an iteration can only lead to $L_{bulk}$ and \textit{not} to $L_{sur}$ or to $L_{EH}$. The claim, sometimes made in the literature, that 
 the Einstein-Hilbert action can be obtained  by starting from the action functional for spin-2 graviton, coupling it to its own stress tensor and iterating the process to all orders, is incorrect (for more details, see \refcite{gravitonmyth}).
Also note that the surface term at linear order
\begin{equation}
\sqrt{-g}L_{sur}\approx\frac{1}{2l_P}\partial_a\partial_b[h^{ab}-\eta^{ab}h^i_i]
\end{equation}
is invariant under the linear gauge transformations $h_{ab}\to h_{ab}+\partial_a\xi_b+\partial_b\xi_a$.  However, the exact form of $L_{sur}$ in \eq{exactsur}
is \textit{not} generally covariant. It is sometimes (again wrongly) claimed in the literature
that if a term is gauge invariant in the linear order, it will be generally covariant in the exact theory; the $L_{sur}$ is a concrete counterexample.

Incidentally, one might wonder what happens in the semiclassical limit in which the the wave functional $\exp iA_{sur}$ will depend on $A_{sur}$, which is not generally covariant. Notice that the foliation independence of this semiclassical limit is ensured if we demand that 
$
\exp iA_{sur} =\exp 2\pi i n
$.
This immediately leads\cite{apporva} to area quantization law:
$\mathcal{A}_\perp=(8\pi L_P^2) n
$. Such results have been around for some time now in different approaches to quantum gravity in which Planck length arises as the lower bound to proper length scales\cite{zeropoint}.
The entire situation is analogous to gauge theory in which `small' gauge transformations leave the action invariant but the `large' gauge transformations do not. One uses a similar quantisation condition on the semiclassical action in that context as well.

\section{Holographic structure of Semiclassical Action for Gravity}
\label{sss:gb}
There is actually a deep reason as to why this works, which  goes beyond the Einstein's theory. Similar results exists for \textit{a wide class of covariant theories based on principle of equivalence}, in which the gravity is described by
a metric tensor $g_{ab}$. Let me briefly describe the general setting from which this thermodynamic picture arises.\cite{cc1}

 Consider a  (generalized) theory of gravity in D-dimensions based on a generally covariant
scalar Lagrangian $L$ which is a functional of the metric $g^{ab}$ and curvature $R^a_{\phantom{a}bcd}$.
Instead of treating $[g^{ab},\partial_cg^{ab},
\partial_d\partial_cg^{ab}$] as the independent variables,  it is convenient to use $[g^{ab},
\Gamma^i_{kl},R^a_{\phantom{a}bcd}]$ as the independent variables. The curvature tensor $R^a_{\phantom{a}bcd}$ can be expressed entirely in terms of $\Gamma^i_{kl}$ and $\partial_j\Gamma^i_{kl}$ and is \textit{independent} of $g^{ab}$. 
To  investigate the general (``off-shell") structure of the theory, let us note that any scalar which depends on  $R^a_{\phantom{a}bcd}$ can be written in the form
$
L=Q_a^{\phantom{a}bcd}R^a_{\phantom{a}bcd}
$ with
the tensor $Q_a^{\phantom{a}bcd}$ depending on curvature and metric. (For example,   any function $L$ can be written as 
$L=Q_a^{\phantom{a}bcd}R^a_{\phantom{a}bcd}$ with $Q_a^{\phantom{a}bcd}=(L/2R)(\delta^c_ag^{bd}-\delta^d_ag^{bc})$
but, of course, other choices of $Q_a^{\phantom{a}bcd}$ can lead to the same $L$.)
We will impose one condition on $Q_a^{\phantom{a}bcd}$ (which can be motivated by considering the variational principle; see ref. \refcite{cc1}); viz.$\nabla_cQ^{ijcd}=0$.
Thus, we shall hereafter confine our attention to Lagrangians of the form:
\begin{equation}
L=Q_a^{\phantom{a}bcd}R^a_{\phantom{a}bcd}; \qquad \nabla_cQ^{ijcd}=0
\end{equation} 
It is easy to show that all these Lagrangian allow the separation:
 \begin{equation}
\sqrt{-g}L=2\sqrt{-g}Q_a^{\phantom{a}bcd}\Gamma^a_{dk}\Gamma^k_{bc}+2\partial_c\left[\sqrt{-g}Q_a^{\phantom{a}bcd}\Gamma^a_{bd}\right]
\equiv L_{\rm bulk} +L_{\rm sur}   
\label{gensq}
 \end{equation} 
with
\begin{equation}
L_{\rm bulk}=2\sqrt{-g}Q_a^{\phantom{a}bcd}\Gamma^a_{dk}\Gamma^k_{bc};\qquad
L_{\rm sur}=2\partial_c\left[\sqrt{-g}Q_a^{\phantom{a}bcd}\Gamma^a_{bd}\right]
\equiv\partial_c\left[\sqrt{-g}V^c\right]
\label{sep}
\end{equation}  
where the last equality defines the D-component object $V^c$, which --- of course --- is not a vector.  The holographic relation between surface and null actions extend to all these cases.
 For example, all these lagrangians satisfy \cite{ps1} a relation of the form:
\begin{equation}
L_{\rm sur} = \partial_p \left(\delta^q_r \frac{\partial L_{\rm bulk}}{\partial \Gamma^q_{pr}}
\right)
\label{holo}
\end{equation} 
 We can also prove the following results \cite{cc1} which determine the bulk and total lagrangians in terms of the surface term (which is probably truer to the spirit of the term holography):
\begin{equation}
L=\frac{1}{2}R^a_{\phantom{a}bcd}\left(\frac{\partial V^c}{\partial \Gamma^a_{bd}}\right);
\quad
L_{bulk}=\sqrt{-g}\left(\frac{\partial V^c}{\partial \Gamma^a_{bd}}\right)
\Gamma^a_{dk}\Gamma^k_{bc}
\label{realholo}
\end{equation}  
Thus the knowledge of the functional form of $L_{sur}$ or --- equivalently --- that of $V^c$
allows us to determine $L_{bulk}$ and even $L$.
The first relation also shows that  $(\partial V^c/\partial \Gamma^a_{bd})$ is generally covariant in spite of the appearance.

More geometrically, writing
 $\mathcal{R}^a_{\phantom{a}b} = (1/2!) R^a_{\phantom{a}bcd} \, w^c \wedge \, w^d$ in terms of the  basis one forms $w^a$,
 introducing a corresponding  form for $Q_{abcd}$ with $\mathcal{Q}^a_{\phantom{a}b} =(1/2!) Q^a_{\phantom{a}bcd}\, w^c \wedge w^d$
 and using $\mathcal{R}^a_{\phantom{a}b} =  d\,   \Gamma^a_{\phantom{a}b} + \Gamma^a_{\phantom{a}c}\, \wedge \, \Gamma^c_{\phantom{c}b}$ where $\Gamma^a_{\phantom{a}b}$ are the curvature  forms,
 we can write the 
 Lagrangian as
 \begin{eqnarray}
(1/2)L &=& \ast \mathcal{Q}^a_{\phantom{a}b}\wedge \mathcal{R}^b_{\phantom{b}a} 
= \ast \mathcal{Q}^a_{\phantom{a}b}\wedge \left( d \Gamma^b_{\phantom{b}a} +  \Gamma^b_{\phantom{b}c}\wedge
\Gamma^c_{\phantom{c}a}\right) \nonumber\\
&=& d\left( \ast \mathcal{Q}^a_{\phantom{a}b}\wedge \Gamma^b_{\phantom{b}a}
\right) + \ast\mathcal{Q}^a_{\phantom{a}b}\wedge \Gamma^b_{\phantom{b}c}\wedge \Gamma^c_{\phantom{c}a}
\end{eqnarray}
provided the $\mathcal{Q}^a_{\phantom{a}b}$ satisfies the condition:
$d\left( \ast \mathcal{Q}^a_{\phantom{a}b}\right) =0$ corresponding to $\nabla_c Q_a^{\phantom{a}bcd} =0$. The separation between bulk and surface terms is obvious.
 We saw earlier  (see \eq{surbulkrel}) that the bulk and surface terms of Einstein-Hilbert action
are related
by a similar identity.
The current result shows that this is a very general result and is based only on the condition $\nabla_a Q^{abcd}=0$.

Everything else goes through as before  and it is possible
to reformulate the theory retaining \textit{only} the surface term for the gravity sector as in the
case of Einstein gravity. (For a related alternative approach, see ref.\refcite{thomas}).
If the original equations of motion of the theory, obtained by standard procedure is $E_{ab}=T_{ab}$, we will now get the equations of the form
$(E_{ab}-T_{ab})\xi^b\xi^a=0$ where $\xi^a$ is \textit{null} if we use appropriate surface term and boundary conditions.  The addition of a cosmological constant
--- by the change $T_{ab}\to T_{ab}+\Lambda g_{ab}$ --- again leaves the equations invariant. 
When combined with  identity $\nabla_a E^{ab}=0$ this will lead to standard field equations with a cosmological term arising as an integration constant: $E_{ab}=T_{ab}+\Lambda g_{ab}$ just as in the case of Einstein-Hilbert action \cite{newper}.
Since the cosmological constant arises only as an integration constant,
 it can be set to any value as a feature of the \textit{solution} to the
field equations in a given physical context.   This provides a basic reason for ignoring the bulk \cc\ 
\textit{and its changes} during various phase transitions in the universe. (Incidentally, the definition for entropy of null vectors, given in
in the first line of \eq{freeenergy} holds even with a more general $Q^{ab}_{cd}$ introduced in this section and one can again obtain the corresponding field equations by varying the null vectors. This approach will be discussed elsewhere.)

What actually determines the specific numerical value of the \cc\ in our universe is a separate question which I will not address here. I have shown elsewhere that,
when coupled to the  thermodynamic paradigm, which suggests  that
 in the presence of a horizon we should  work with the degrees of freedom confined by the horizon, it is possible to predict the value of this integration constant\cite{cc1,tpcqglamda}.

\subsection{Structure of low-energy effective action for gravity} 
 
Let us now consider the explicit form of divergence-free fourth rank tensor $Q_a^{\phantom{a}bcd}$, having the symmetries of the curvature tensor, 
 which determines the structure of the theory. The semiclassical,
low energy, action for gravity can now be determined from the derivative expansion
of $Q_a^{\phantom{a}bcd}$ in powers of number of derivatives: 
\begin{equation}
Q_a^{\phantom{a}bcd} (g,R) = \overset{(0)}{Q}_a{}^{bcd} (g) + \alpha\, \overset{(1)}{Q}_a{}^{bcd} (g,R) + \beta\, \overset{(2)}{Q}_a{}^{bcd} (g,R,\nabla R) + \cdots
\label{derexp}
\end{equation} 
where $\alpha, \beta, \cdots$ are coupling constants. We will treat the expansion in terms of the number of derivatives as giving the quantum corrections to the classical theory. To determine the first term, say, we only need to obtain all the possible fourth rank tensors $Q^{abcd}$ which (i) have the symmetries of curvature tensor; (b) are divergence-free and (iii) are made from $g^{ab}$; similarly, to obtain the next term, we allow the tensor $Q^{abcd}$ to depend on $g^{ab}$ and  $R^a_{\phantom{a}bcd}$ etc. 
Interestingly enough, at the first two orders, this leads to \textit{all} the gravitational theories (in D dimensions) in which
the field equations are no higher than second degree, though we did \textit{not }demand that explicitly. At the lowest order, 
if we do not use the curvature tensor, then we have just one unique choice for zeroth order,
 made from metric:
$
Q_a^{(0)bcd}=(1/2)(\delta^c_ag^{bd}-\delta^d_ag^{bc})
$ 
which satisfies our constraints leading to the standard Einstein's theory\cite{ps2}.

Next, if we allow for $Q_a^{\phantom{a}bcd}$ to depend linearly on curvature, then we have the following 
additional choice of  tensor with required symmetries:
\begin{equation}
\overset{(1)}{Q}_{abcd}=R_{abcd} -  G_{ac}g_{bd}+ G_{bc}g_{ad} + R_{ad}g_{bc} - R_{bd}g_{ac} 
\label{ping}
\end{equation} 
(In four dimensions, this tensor is essentially the double-dual of
$R_{abcd}$ and in any dimension can be obtained from $R_{abcd}$ using the alternating tensor \cite{love}.)
In this case,  we get
\begin{eqnarray}
L&=&\frac{1}{2}\left(g_{ia}g^{bj}g^{ck}g^{dl}-4g_{ia}g^{bd}g^{ck}g^{jl}
+\delta^c_a\delta^k_ig^{bd}g^{jl}\right)R^i_{\phantom{i}jkl}R^a_{\phantom{a}bcd}\nonumber\\
&=&\frac{1}{2}\left[R^{abcd}R_{abcd}-4R^{ab}R_{ab}+R^2\right]
\end{eqnarray} 
This is just the Gauss-Bonnet(GB) action which is a pure divergence in 4 dimensions but not in higher dimensions. The fact\cite{zw} that \textit{string theoretical models get GB type terms as corrections} is noteworthy in this regard. 
We can similarly determine the higher order corrections. 
Both  Einstein-Hilbert lagrangian and  Gauss-Bonnet  lagrangian can be written in a condensed notation using alternating tensors as: 
\begin{equation}
L_{EH}=\delta^{13}_{24}R^{24}_{13};\qquad 
L_{GB}=\delta^{1357}_{2468}R^{24}_{13}R^{68}_{57}
\end{equation} 
where the numeral $n$ actually stands for an index $a_n$ etc. The obvious generalization leads to the Lanczos-Lovelock lagrangian\cite{love}:
\begin{equation}
L_m=\delta^{1357...2k-1}_{2468...2k}R^{24}_{13}R^{68}_{57}
....R^{2k-2\,2k}_{2k-3\,2k-1}; \qquad k=2m
\label{lll}
\end{equation} 
where $k=2m$ is an even number. The $L_m$ is clearly a homogeneous function of degree $m$ in
the curvature tensor $R^{ab}_{cd}$. In this case, 
 $Q_a^{\phantom{a}bcd}$ is an $n$th order polynomial in the curvature tensor. All these actions obey the holographic relation between the surface and bulk terms, generalizing the \eq{surbulkrel} of Einstein-Hilbert action. It can be shown that\cite{ayan}, in general, 
\begin{equation}
[(D/2) - m]L_{sur} =-\partial_i \left[ g_{ab} \frac{\delta L_{bulk}}{\delta (\partial_i g_{ab})}
  +\partial_jg_{ab} \frac{\partial L_{bulk}}{\partial (\partial_i \partial_jg_{ab})}
  \right] 
\label{result1} 
\end{equation} 
which again has a nice `$d(pq)$ structure' as in the case of Einstein-Hilbert action.
Also note that, as long as the  higher order quantum gravitational corrections are determined by the holographic principles, the higher orders terms will all respect the invariance of the theory under $T_{ab}\to T_{ab}+\lambda g_{ab}$ and the cosmological constant will continue to remain an integration constant even when quantum corrections are incorporated.

\subsection{Surface term as Entropy of horizons}

Finally, it can be shown that the surface term is closely related to the entropy of horizons even in the generalized context. I will now briefly indicate the nature of this proof\cite{ayan}. 
To do this we need an expression for the entropy of the horizon in a general context when the lagrangian depends of $R^a_{\phantom{a}bcd}$ in a non-trivial manner. Such a formula has been provided by Wald in ref. \refcite{noether} and can be expressed as a integral over $P_a^{\phantom{a}bcd}\equiv (\partial L/\partial R^a_{\phantom{a}bcd} )$ on the horizon, evaluated on-shell. It can also been shown\cite{noether} that this definition is equivalent to
interpreting  entropy as the Noether charge associated with diffeomorphism invariance. We shall briefly summarize this approach and use this definition.

To define the Noether charge associated with the diffeomorphism invariance, let us consider the variation $x^a\to x^a + \xi^a$ under which the metric changes by 
$\delta g_{ab} = - (\nabla_a \xi_b + \nabla_b \xi_a)$. The change in the action, when
evaluated on-shell, is contributed only by the surface term so that we have the 
relation
\begin{equation}
\delta_\xi A\big|_{\rm on\ shell} = - \mes{D}{} \nabla_a (L\xi^a) 
                                  = \mes{D}{} \nabla_a(\delta_\xi V^a)
\end{equation} 
[The subscript $\xi$ on $\delta_\xi....$ is a reminder that we are considering the changes due to a particular kind of variation, viz. when the metric changes by $\delta g_{ab} = - (\nabla_a \xi_b + \nabla_b \xi_a)$.]
This leads to the conservation law $\nabla_a J^a =0$ with $J^a = L \xi^a + (\delta_\xi V^a)
\equiv \nabla_b J^{ab}$ with the last equality defining the antisymmetric tensor
$J^{ab}$.
For a lagrangian of the type $L=L(g^{ab},R^a_{bcd})$ direct computation  shows that $J^{ab}$ is given by:
\begin{equation}
J^{ab} = - 2 P^{abcd} \nabla_c \xi_d + 4 \xi_d \left(\nabla_c P^{abcd}\right)
\label{noedef}
\end{equation} 
with $P_{abcd}\equiv (\partial L/\partial R^{abcd})$.
We shall  confine ourselves to  Lanczos-Lovelock type lagrangians for which 
 \begin{equation}
L = \frac{1}{m} R^{abcd} \left( \frac{\partial L}{\partial R^{abcd}}\right) \equiv R^{abcd} Q_{abcd}
\end{equation} 
with $\nabla_a P^{abcd} =\nabla_a Q^{abcd} =0$ so that $J^{ab} = - 2 P^{abcd} \nabla_c \xi_d$.

We want to evaluate the Noether charge corresponding to the current $J^a$ for a static metric with a bifurcation horizon and a killing vector field $\xi^a = (1, \textbf{0})$. The location of the horizon is given by the vanishing of the norm $\xi^a\xi_a = g_{00},$ of this  killing vector.
Using these facts as well as the relations  $\nabla_c\xi^d = \Gamma^d_{c0}$ etc., we find that
$J^{ab} = 2P_{d}^{\phantom{d}0ab}\Gamma^d_{c0}$. Therefore the  Noether charge is given by
\begin{equation}
\mathcal{N} = \mes{D-1}{t} J^0 
=\mes{D-2}{t,r_H} J^{r0}
\end{equation} 
in which we have ignored the contributions arising from $b=$ transverse directions. This is
justifiable when transverse directions are compact or in the case of Rindler approximation
when nothing changes along the transverse direction. In the radial direction, we have again confined to the contribution at $r=r_H$ which is taken to be the location of the horizon.
Using 
$Q^{r0} = 2 P^{dcr0} \Gamma_{dc0} = -2 P^{dcr0}\partial_d g_{c0}$ we get
\begin{equation}
\mathcal{N} = -2 \mes{D-2}{t,r_H} P^{dcr0}\partial_d g_{c0} = 2m \mes{D-2}{t,r_H} Q^{cdr0}\partial_d g_{c0}
\label{n}
\end{equation} 
The dimension of $\mathcal{N}$ is $L^{D-3}$ which is area of transverse dimensions
divided by a length. Entropy, which has the dimensions of transverse area, is given by the product of $\mathcal{N}$ and the interval in time integration. If the surface gravity of the horizon is $\kappa$, the time integration can be limited to the range $(0,\beta)$ where $\beta=2\pi/\kappa$. The entropy, computed from the Noether charge approach is thus given by
\begin{equation}
S_{Noether}=\beta\mathcal{N} = 2\beta m \mes{D-2}{t,r_H} Q^{cdr0}\partial_d g_{c0}
\label{s}
\end{equation}  

We will now show that this is the same  result one obtains by evaluating our surface term on the horizon except for a proportionality constant. In the stationary case, the contribution of surface term on the horizon is given by
\begin{equation}
S_{sur} = 2 \int d^D x \, \partial_c \left[ \sqrt{-g} Q^{abcd} \partial_b g_{ad}\right]
= 2\int dt\mes{D-2}{r_H} Q^{abrd} \partial_b g_{ad}
\end{equation} 
Once again, taking the integration along $t$ to be in the range $(0,\beta)$ and ignoring
transverse directions, we get
\begin{equation}
S_{sur} = 2\beta \mes{D-2}{r_H} Q^{abr0} \partial_b g_{a0}
\end{equation} 
Comparing with \eq{n}, we find that 
\begin{equation}
S_{Noether} = mS_{sur}
\end{equation} 
The overall proportionality factor has a simple physical meaning. It can be shown that\cite{ayan}  the quantity $mL_{sur}$, rather than $L_{sur}$, which has the $``d(qp)"$ structure and it is this particular combination which plays the role of entropy, as to be expected\cite{teitel}.

The interpretation finds additional strength from the following fact: We saw earlier that,
in the case of Einstein-Hilbert gravity, it is possible to
interpret Einstein's equations as the thermodynamic identity $TdS = dE
+ PdV$ for a spherically symmetric spacetime and thus provide a 
thermodynamic route to understand the dynamics of gravity. It can be shown\cite{aseem} that  the field equations
for Lanczos-Lovelock action can  also be expressed as $TdS = dE + PdV$ with $S$ and $E$
being given by expressions previously derived in the literature by
other approaches. This  result
indicates a deep connection between the thermodynamics of horizons and
the allowed quantum corrections to standard Einstein gravity, and
shows that the relation $TdS = dE + PdV$ has a greater domain of
validity that Einstein's field equations.

 \section{Conclusions}
 
 The approach highlights the role of null surfaces --- which block information and act as horizons for a congruence of observers locally --- in the formulation of the theory. The intriguing analogy between the gravitational dynamics of
horizons  and thermodynamics  is not yet understood at a deeper
level. One possible way of interpreting these results is
to assume that spacetime is analogous to an elastic solid and
equations describing its dynamics are similar to those of elasticity,
(the ``Sakharov paradigm"; see e.g., ref. \refcite{sakharov}). The unknown,
microscopic degrees of freedom of spacetime (which should be analogous
to the atoms in the case of solids) is normally expected to  play a role only when
spacetime is probed at Planck scales (which would be analogous to the
lattice spacing of a solid). The exception to this general rule arises 
when we consider horizons \cite{magglass} which have finite
temperature and block information from a family of observers. In a
manner which is not fully understood, the horizons link certain
aspects of microscopic physics with  the bulk dynamics just as
thermodynamics can provide a link between statistical mechanics and
(zero temperature) dynamics of a solid.

 In this approach, the full theory has some microscopic variables $q_i$ and an  action $A_{micro}(q_i)$. Integrating out short wavelength fluctuations and microscopic degrees of freedom  should lead to a  long wavelength effective action, which could be a pure surface term, 
 as well as bring about $g_{ab}$ as the new dynamical variables in terms of which the effective action is described. 
 This will lead to \textit{the effective low energy  degrees of freedom of gravity
for a volume $\mathcal{V}$ to
reside in its boundary $\partial\mathcal{V}$} --- a  point of view that is strongly supported by the study
of horizon entropy, which shows that the degrees of freedom hidden by a horizon scales as the area and not as the volume.

 In this description $g_{ab}$ is like the density $\rho$ of a solid arising from large number of atoms and is not a fundamental dynamical variable. It does not make sense to vary the
 $g_{ab}$ \textit{arbitrarily} in this action.
 Instead, the (covariant) equations of motion are obtained by demanding the 
invariance of the (noncovariant) surface action $A_{sur}$,
 under virtual displacements of any (observer dependent) horizon normal to itself.
 This might seem unusual at first but, as I explained before, it arises from the thermodynamic interpretation of (observer dependent) horizons.  In the displacement $x^a\to \bar{x}^a=x^a+\xi^a$ the
$\xi^a(x)$ is similar to the displacement vector field used, for example,
in the study of elastic solids. 
The true degrees of freedom are some unknown `atoms of spacetime' but in the continuum limit,
the displacement $x^a\to \bar{x}^a=x^a+\xi^a(x)$ captures the relevant dynamics,  just like 
in the study of elastic properties of the continuum solid. In fact, one can reformulate the Einstein gravity in terms of
the dynamics of a null vector field in a background spacetime\cite{elastic}.
The horizons in the spacetime are then  similar to defects in the solid so that their displacement costs entropy. 

The approach also leads to two new insights, which one could not have been anticipated a priori. 
First, the surface action leads in a natural fashion to equations
$(E_{ab}-T_{ab})\xi^a\xi^b=0$ for all null vectors $\xi^a.$  These equations of motion are now invariant
under the changes to the vacuum energy $T_{ab}\to T_{ab}+\Lambda g_{ab}$ and we have a natural solution to the \cc\ problem.
Note that, this approach, unlike many others, can handle the \textit{changes} to the vacuum energy density arising due to phase transitions in the early universe. The observed \cc\ can be interpreted\cite{cc2}  as arising due to the vacuum fluctuations in a region confined by the horizon and --- in that sense --- is coupled to the surface degrees of freedom of gravity. 

Second the effective action in  can be expanded in terms of number of derivatives and the
low energy effective action for gravity is then determined by the derivative expansion
of $Q_a^{\phantom{a}bcd}$ in powers of number of derivatives, given by \eq{derexp}:
The first term leads to Einstein-Hilbert action and the second one leads to the Gauss-Bonnet action; this (as well as higher order terms) have a natural interpretation of being a quantum correction in this approach. 
We also have a general principle for determining
the correction terms (by constructing the divergence free tensor $Q_a^{\phantom{a}bcd}$ from variables with right number of derivatives) and constraining the structure of underlying theory. 
It is worth recalling \textit{that such a Gauss-Bonnet term
arises as the correction in string theories}\cite{zw}. The thermodynamic interpretation (which is on-shell) as well as
the holographic description (which is off-shell) are also applicable to quantum corrections
to the Einstein-Hilbert Lagrangian. 
The invariance of the theory under $T_{ab}\to T_{ab}+\Lambda g_{ab}$ continues to hold for the higher order terms as well suggesting that the mechanism for ignoring the bulk \cc\ is likely to survive quantum gravitational corrections.

\section*{Acknowledgements} I thank M. Charest, N. Dadhich, S. Guo, S. Liberati,   A. Paranjape, K. Subramanian and H. Zhang for useful correspondence and discussions.

\section*{Appendix: Surface Term in Einstein-Hilbert action}

The Einstein-Hilbert action for gravity in 4-dimensions is given by (we use signature $-+++$
and the convention $R^i_{\phantom{i}jkl}=\partial_k\Gamma_{lj}^i-\partial_l\Gamma_{kj}^i+\cdots$;
Latin letters range over 0-3 and Greek letters range over 1-3):
\begin{equation}
16\pi A_{EH}= \int_\mathcal{V} d^4x\sqrt{-g}\, L_{EH} =\int_\mathcal{V} d^4x\sqrt{-g}\, R
\end{equation} 
Since $L_{EH}$  is linear in second derivatives of the metric, it is clear that $\sqrt{-g}L_{EH}$ can be written as  $\sqrt{-g}L_{EH}=\sqrt{-g}L_{bulk}-L_{sur}$ where $L_{bulk}$ is quadratic in the first derivatives of the metric and
$ L_{sur}$ is a total derivative which leads to a surface term in the action.
Explicitly:
  \begin{equation}
  R\sqrt{-g} = \sqrt{-g}L_{bulk}-L_{sur}\equiv \sqrt{-g} L_{bulk}  - \partial_j P^j
  \label{eqnone}
  \end{equation}
where   
  \begin{equation}
    L_{\rm bulk} = g^{ab} \left(\Gamma^i_{ja} \Gamma^j_{ib} -\Gamma^i_{ab} \Gamma^j_{ij}\right)
    \label{lquad}
    \end{equation}
      and 
    \begin{equation}
    P^c \equiv\sqrt{-g}V^c= \sqrt{-g} \left(g^{ck} \Gamma^m_{km}- g^{ik} \Gamma^c_{ik}\right) =-
     \frac{1}{\sqrt{-g}} \partial_b (g g^{bc})
    \label{defpcone}
     \end{equation}
where the first equality defines the 4-component object $V^c=g^{-1}\partial_b[gg^{bc}]$, which --- of course --- is not a vector. We will also use the \textit{notation} of covariant derivative operator $\nabla_i=(\sqrt{-g})^{-1}\partial_i(\sqrt{-g}....)$ and write
\begin{equation}
L_{sur}=\partial_c P^c=\partial_c(\sqrt{-g}V^c)=\sqrt{-g}\nabla_c V^c
\end{equation} 
with the clear understanding that $V^c$ is not a vector.

Integrating $L_{sur}$ over a four-dimensional region leads to the surface term in the action:
\begin{equation}
16\pi A_{sur}=
\int_{\mathcal{V}} d^4 x L_{sur}=\int_{\mathcal{V}} d^4 x\sqrt{-g}\nabla_c V^c=
\int_{\partial\mathcal{V}} d^3 x  
\sqrt{h}n_cV^c
\label{actfunc}
\end{equation}
where $n_c$ is the unit normal on $\partial\mathcal{V}$ and $h$ is the determinant
of the metric induced on the surface. (Little thought --- or simple algebra --- shows that the  Gauss theorem  holds even when  $V^c$ not a vector, as long as computations are carried out in a specific coordinate system). Obviously, $A_{sur}$ is {\it not} generally covariant.
This term, $A_{\rm sur}$, when added to Einstein-Hilbert action leads to an action which is purely quadratic but will \textit{not} be generally covariant.

Before proceeding further, I want to briefly compare $A_{sur}$ with another surface action\cite{gh} involving the trace of the extrinsic curvature:

\begin{equation}
16\pi A_{sur}^{GH}=2\int_{\partial\mathcal{V}} d^3 x  \sqrt{h}K =2\int_{\mathcal{V}} d^4 x\sqrt{-g}\nabla_c (Kn^c)
\label{actiondiff}
\end{equation} 
In the second relation we have extended the vector field $n_c$ --- originally defined only on $\partial\mathcal{V}$ --- in any arbitrary fashion into the bulk $\mathcal{V}$
and taken $K = - \nabla_i n^i$. This is legal because only its value on $\partial\mathcal{V}$ contributes to the integral.
The difference between the two actions $A_{sur}$ and $A_{sur}^{GH}$   is
\begin{equation}
16\pi(A_{sur}-A_{sur}^{GH})=\int_{\mathcal{V}} d^4 x\partial_c[P^c-2\sqrt{-g}Kn^c]
\end{equation} 
which can be easily computed. As an example, let us consider the contributions from the $\partial\mathcal{V}$ that is made of $t=$ constant surface. Then, for the metric, parameterized as
\begin{equation}
ds^2=-(Ndt)^2+h_{\mu\nu}[dx^\mu +N^\mu dt][dx^\nu +N^\nu dt]
\end{equation} 
[so that $g^{00}=-1/N^2,g^{0\mu}=N^\mu/N^2,g^{\mu\nu}=h^{\mu\nu}-N^\mu N^\nu/N^2, g=-N^2h$] we can  compute the difference to be
\begin{equation}
16\pi(A_{sur}-A_{sur}^{GH})=\int_{t} d^3 x[P^0-2\sqrt{-g}Kn^0]=-\int_{t} d^3 x\sqrt{h}
\left[\frac{\partial_\mu N^\mu}{N}\right]
\end{equation} 
This result shows several important features. 
\begin{itemize}
\item 
To begin with $(A_{sur}-A_{sur}^{GH})$ is in general \textit{nonzero}. If the coordinates are chosen such that the surface $\partial\mathcal{V}$ corresponds to  $x^M=$ constant, then ($A_{sur}-A_{sur}^{GH})=0$ only for the coordinate choice in which the metric has no off-diagonal terms with respect to the coordinate labelled by $M$; in the case of constant time foliation, this requires $g^{0\mu}=0$. The absolute value of the action has some significance in semiclassical gravity etc. so it is important to note that $A_{sur}\neq A_{sur}^{GH}$ in general.
\item  
Suppose one is interested in the \textit{variation} of these 
actions (rather than their values) when the metric is varied arbitrarily. Again, in general, $(\delta A_{sur}-\delta A_{sur}^{GH})\neq0$. But if we consider variations with $g^{ab}$ held fixed on  $\partial\mathcal{V}$, then $N$ and $h$ will not vary on  $\partial\mathcal{V}$; further, if the metric is fixed everywhere on $\partial\mathcal{V}$, then the \textit{spatial} derivative
$\partial_\mu N^\mu$ is also fixed everywhere on $\partial\mathcal{V}$ and cannot contribute to the variation. So we find that $\delta A_{sur}=\delta A_{sur}^{GH}$ for variations of metric with
$g^{ab}$ held fixed on  $\partial\mathcal{V}$. This is why either $A_{sur}$ or $A_{sur}^{GH}$
can be added to Hilbert action to obtain a sensible variational principle with $g^{ab}$ held fixed on  $\partial\mathcal{V}$.
\item The quadratic action obtained as $A_{EH}+A_{sur}$ is an integral over $L_{quad}$ in \eq{lquad} and it is obvious that this expression is noncovariant. On, the other hand, the quadratic action $A_{EH}+A_{sur}^{GH}$ can also be expressed as an integral over the \textit{local} lagrangian:
\begin{equation}
L^{GH}_{quad}=R+2\nabla_i(Kn^i)=R-2\nabla_i(n^i\nabla_jn^j)
\end{equation} 
Such a lagrangian can be interpreted as generally covariant but \textit{it is foliation dependent} which is as bad as being non-covariant. 
In fact, almost any noncovariant expression can be written in a generally covariant manner if one is allowed to introduce extra vector field characterizing the foliation. For example, one would have considered a component
of a tensor, say, $T_{00}$ as not generally covariant. But a quantity $\rho=T_{ab}u^au^b$ is a generally covariant scalar which will reduce to $T_{00}$ in a local frame in which $u^a=(1,0,0,0)$.
It is appropriate to say that $\rho$ is generally covariant but foliation dependent. The $A_{GH}$ uses the normal vector $n^i$ of the boundary in a similar manner.
 
\end{itemize}

I will now continue with the discussion of $A_{sur}$ 
and consider a total action for gravity and matter by adding to $A_{sur}$ the matter action; that is, 
\begin{equation}
A_{tot}=A_{sur}+A_{matter}[\phi_i,g]
\end{equation}
where $A_{matter}[\phi_i,g]$ is the standard matter action  in a spacetime with metric
$g_{ab}$. The $\phi_i$ denotes some generic matter degrees of freedom; varying $\phi_i$
will lead to standard equations of motion for matter in a background metric and these equations will also ensure that the energy momentum tensor
of matter $T^a_b$ satisfies $\nabla_aT^a_b=0$.
I want to compute the variation of 
$A_{tot}$ when the metric changes by $g^{ab}\to g^{ab}+\delta g^{ab}$ with all other matter variables remaining unchanged.

The variation of matter part is easy. When the metric changes by
$\delta g^{ab}$ and the matter action changes by
\begin{equation}
\delta A_{matt}=-\frac{1}{2}\int_\mathcal{V}d^4x\sqrt{-g}T_{ab}\delta g^{ab}
\label{delmat}
\end{equation}
To obtain the variation of $A_{sur}$, we will first write $L_{sur}$ in a slightly simplified notation. It is easy to see from \eq{defpcone}
that
\begin{equation}
L_{\rm sur}
=2Q_{ak}^{cd}\partial_c\left[\sqrt{-g}g^{bk}\Gamma^a_{bd}\right]
\label{sep2}
\end{equation} 
where $Q_{ak}^{cd}$ is the alternating (`determinant') tensor:
$
2Q_{ab}^{cd}=(\delta^d_a\delta^c_b-\delta^c_a\delta^d_b)
$
which is, of course,  a constant (The usual symbol is $\delta_{ab}^{cd}$ which we avoid for conflict of notation). Therefore
\begin{equation}
\delta L_{\rm sur}
=2Q_{ak}^{cd}\partial_c\left[
\sqrt{-g}g^{bk}\delta \Gamma^a_{bd} +\Gamma^a_{bd}\delta(\sqrt{-g}g^{bk}) 
\right]
\label{dellsur}
\end{equation}  
Using
\begin{equation}
\delta(\sqrt{-g}g^{bk})
=\sqrt{-g}[\delta_l^b\delta_m^k -\frac{1}{2}g^{bk} g_{lm}]\delta g^{lm}
\equiv \sqrt{-g}B^{bk}_{lm}\delta g^{lm}
\end{equation} 
the second term in \eq{dellsur} can be written as
\begin{eqnarray}
2Q_{ak}^{cd}\partial_c\left[
\Gamma^a_{bd}\delta(\sqrt{-g}g^{bk}) \right]
&=&2Q_{ak}^{cd}\partial_c\left[\sqrt{-g}\Gamma^a_{bd}B^{bk}_{lm}\delta g^{lm}\right]
\equiv\partial_c[\sqrt{-g}M^c_{\phantom{c}lm}\delta g^{lm}]\nonumber\\
&=&\sqrt{-g}\nabla_c[M^c_{\phantom{c}lm}\delta g^{lm}]
\end{eqnarray} 
where we have defined the 3-index \textit{non}tensorial object,
 \begin{equation}
 M^c_{\phantom{c}lm}=2Q_{ak}^{cd}B^{bk}_{lm}\Gamma^a_{bd}=
 -\Gamma^c_{lm}+\Gamma^d_{ld}\delta^c_m-\frac{1}{2}g_{lm}V^c
 \label{mclmeqn}
\end{equation} 
for ease of notation. (Its explicit form is irrelevant  for our discussion).

Similarly the first term in \eq{dellsur} is
\begin{equation}
2Q_{ak}^{cd}\partial_c\left[
\sqrt{-g}g^{bk}\delta \Gamma^a_{bd} \right] =\sqrt{-g}\nabla_c (N_a^{\phantom{a}bdc}\delta\Gamma^a_{bd})
\label{eqfirst}
\end{equation}
with $N_a^{\phantom{a}bdc}\equiv2Q_{ak}^{cd}g^{bk}$ being  a genuine tensor. Combining the two terms and integrating over four volume, we get:
\begin{eqnarray}
16\pi \delta A_{sur}&=& \int_{\mathcal{V}} d^4 x\,  \sqrt{-g}\nabla_c (N_a^{\phantom{a}bdc}\delta\Gamma^a_{bd})+
\int_{\mathcal{V}} d^4 x\,\sqrt{-g}\nabla_c[M^c_{\phantom{c}lm}\delta g^{lm}]\nonumber\\
&=&\int_{\partial\mathcal{V}} d^3 x\,  \sqrt{h}(n_c N_a^{\phantom{a}bdc})\delta\Gamma^a_{bd}+
\int_{\partial\mathcal{V}} d^3 x\,\sqrt{h}\,(n_cM^c_{\phantom{c}lm})\delta g^{lm}
\label{fourvol}
\end{eqnarray} 
Note that the variation of the surface action has two pieces depending on $\delta\Gamma^a_{bd}$ and on $\delta g^{lm}$. For a completely arbitrary variation of the metric both the terms will contribute. However if we consider variations of the type with $\delta g^{lm}=0$
on $\partial\mathcal{V}$, the second term will not contribute. In the first term, derivatives of the metric taken along the surface will have
zero variation; but  $\delta\Gamma^a_{bd}$ will have contributions from the variation of  derivatives of the metric normal to the surface. Thus
even when $\delta g^{lm}=0$
on $\partial\mathcal{V}$, the $\delta A_{sur}$ does \textit{not} vanish.
(Of course, this is root cause of the standard problem with Einstein-Hilbert action; just keeping $\delta g^{lm}=0$
on $\partial\mathcal{V}$ does not lead to field equations because of the nonzero first term --- which is just the standard result in fancy notation). I will now put this fact to good use.
 
To see this explicitly and to proceed further, it is better to write the first term in \eq{fourvol} in a different manner and keep it as an integral over the bulk region $\mathcal{V}$ for the moment.
Using the fact that $\delta\Gamma$'s are tensors and working in the local inertial frame, we can
also express this variation in \eq{eqfirst}  in the form of  the standard textbook result:
\begin{equation}
2Q_{ak}^{cd}\partial_c\left[
\sqrt{-g}g^{bk}\delta \Gamma^a_{bd} \right] =-\sqrt{-g}g^{ab}\delta R_{ab}
\end{equation}
This allows us to write:
\begin{equation}
16\pi \delta A_{sur}
=-\int_{\mathcal{V}} d^4 x\,  \sqrt{-g}g^{ab}\delta R_{ab}+
\int_{\partial\mathcal{V}} d^3 x\sqrt{h}\,(n_c M^c_{\phantom{c}lm})\delta g^{lm}
\label{mattvar}
\end{equation} 
Adding the matter variation, we get
\begin{eqnarray}
16\pi \delta A_{tot}&=& -\int_{\mathcal{V}} d^4 x\,  \sqrt{-g}g^{ab}\delta R_{ab}+
\int_{\partial\mathcal{V}} d^3 x\sqrt{h}\,(n_c M^c_{\phantom{c}lm})\delta g^{lm}\nonumber\\
&&\qquad -8\pi\int_\mathcal{V}d^4x\sqrt{-g}T_{ab}\delta g^{ab}
\label{mattvartwo}
\end{eqnarray}

Up to this point everything has been quite general and we did not assume anything about the form of $\delta g^{ab}$. We will now specialize to 
 variations of the metric of the form $\delta g^{ab}=\nabla^a\xi^b+\nabla^b\xi^a$
where $\xi^a$ is (at present) an unspecified vector field.
The change in the matter action is now:
\begin{equation}
\delta A_{matt}=-\frac{1}{2}\int_\mathcal{V}d^4x\sqrt{-g}T_{ab}\delta g^{ab}
=-\int_\mathcal{V}d^4x\sqrt{-g}\nabla_a(T^a_b\xi^b)
\label{delmat1}
\end{equation}
where we have used the fact that $\nabla_a T^a_b=0$, which arises from the  equations of motion for the matter. 
As discussed in the main text,
we are \textit{not} introducing a coordinate shift $x^a\to \bar{x}^a=x^a+\xi^a$ but merely choosing a a specific type of $\delta g^{ab}$ parametrized by a vector field $\xi^a$; this is why matter fields are not varied. However, this
distinction is not relevant in the gravity sector for variables which depends \textit{only} on metric and its derivatives. Changes in $g_{ab},\Gamma^i_{jk}$, their derivatives etc. induced under a coordinate shift $x^a\to \bar{x}^a=x^a+\xi^a$ will be identical to those calculated using $\delta g^{ab}=\nabla^a\xi^b+\nabla^b\xi^a$ in the expressions. But if one is using expressions in action involving other vector fields, like the normal $n_i$ to $\partial\mathcal{V}$
(as in $A^{GH}_{sur}$ of \eq{actiondiff}), then again the resulting variations will be different if we just change the metric by $\delta g^{ab}=\nabla^a\xi^b+\nabla^b\xi^a$ or interpret it as arising from a coordinate transformation
$x^a\to \bar{x}^a=x^a+\xi^a$. In the former case, other vector fields like $n_i$ do not change
(that is, they do not change intrinsically; of course if $n_i$ does not change, $n^i$ will reflect the change in metric) but if coordinates are transformed these vector fields need to change. It follows that an expression which is foliation dependent will transform differently in these two case. We are not talking about diffeomorphism.

To proceed further, we need to know the variation $g^{ab}\delta R_{ab}$ under the transformation
 $\delta g^{ab}=\nabla^a\xi^b+\nabla^b\xi^a$. The result
 is:
\begin{equation}
\sqrt{-g} g^{ab}\delta R_{ab}=-2\sqrt{-g}\nabla^a(R_{ab}\xi^b) 
\label{twofour}
\end{equation} 
To prove \eq{twofour} we will start with the result (which can be easily proved
by expressing $\delta \Gamma^a_{bc}$ in terms of $\delta g^{ij}$)
\begin{equation}
g^{ik}\delta R_{ik} = \nabla_a\nabla^a (g_{ik}\delta g^{ik}) - \nabla_a\nabla_b (\delta g^{ab})
\end{equation} 
When $\delta g^{ab} = \nabla^a\xi^b + \nabla^b\xi^a$, this expression becomes
\begin{equation}
g^{ik}\delta R_{ik} = \nabla_a\left( 2 \nabla^a \nabla_j \xi^i - \nabla_b \nabla^a \xi^b -
\nabla_b \nabla^b \xi^a\right)
\end{equation} 
In the middle term, we  write $\nabla_b \nabla^a \xi^b = \nabla^a \nabla_b \xi^b + R^a_i \xi^i$
thereby getting
\begin{equation}
g^{ik}\delta R_{ik} = -\nabla_a (R^a_i \xi^i) + \nabla_a\left( 2 \nabla^a \nabla_j \xi^i   -
\nabla_b \nabla^b \xi^a\right)
\end{equation} 
Next, we note that $\nabla_a\nabla_b \nabla^b\xi^a = \nabla_b\nabla_a \nabla^b\xi^a$.
(This is just a special case of the, easily proved, general result $\nabla_a\nabla_b H^{ba} = \nabla_b\nabla_a H^{ba}$ for any tensor $H^{ab}$)
Finally, expressing $\nabla_a \nabla^b\xi^a$ in terms of $\nabla^b \nabla_a\xi^a$
we find that one more term cancels leading to the result
\begin{equation}
g^{ik}\delta R_{ik} = -2 \nabla_b ( R^b_j \xi^j)
\end{equation} 

(I went through this `first principle' proof just to illustrate that the result is purely an algebraic consequence of $\delta g^{ab} = \nabla^a\xi^b + \nabla^b\xi^a$ and it has nothing to do with coordinate transformations. There is, of course, a much shorter route to the result by using 
a trick\cite{newper}.
  We know that $\delta R=g^{ik}\delta R_{ik}+R_{ik}\delta g^{ik}$ so that
\begin{equation}
g^{ik}\delta R_{ik}=\delta R -R_{ik}\delta g^{ik} =\delta R -2R_{ik}\nabla^i\xi^k
\label{ooo}
\end{equation} 
The determine the form of $\delta R$ we argue as follows: We know that, under the infinitesimal
coordinate transformation, $x^a\to \bar{x}^a=x^a+\xi^a$, we have $\delta R=-\xi^a\partial_aR$
and $\delta g^{ab} = \nabla^a\xi^b + \nabla^b\xi^a$. But since one can always determine $\delta R$
entirely in terms of $\delta g^{ab}$, it follows that $\delta R=-\xi^a\partial_aR$ if
$\delta g^{ab} = \nabla^a\xi^b + \nabla^b\xi^a$, \textit{irrespective whether we make any  coordinate transformation or not}. Substituting into \eq{ooo}, we get
\begin{eqnarray}
g^{ik}\delta R_{ik}&=& -\xi^k\partial_kR  -2R_{ik}\nabla^i\xi^k\nonumber \\
&=&-2\nabla^i (R_{ik}\xi^k) +2\xi^k\nabla^i(R_{ik}-\frac{1}{2}g_{ik}R)=-2\nabla^i (R_{ik}\xi^k)
\end{eqnarray} 
since the second term vanishes due to Bianchi identity. 
 The use of $x^a\to \bar{x}^a=x^a+\xi^a$ to determine the form of $\delta R$ is simply a trick to save algebra and should not be confused with genuine coordinate transformations.)

Substituting these into \eq{mattvartwo}  we get the final result  we needed in the text:
\begin{equation}
16\pi\delta A_{tot}
=2\int_{\partial\mathcal{V}}d^{3}x\sqrt{h}
(R^a_b-8\pi T^a_b)\xi^bn_a
-\int_{\partial\mathcal{V}}d^{3}x\, \sqrt{h}\, (n_cM^c_{\phantom{c}lm}\delta g^{lm})
\label{appdeltot}
\end{equation}
The rest of the analysis is described in the main text.


\begin{thebibliography}{000}
\bibitem{ps1}
In evaluating the partial derivative on the right hand side we hold $Q_a^{\phantom{a}bcd}$
fixed and treat all components of $\Gamma^k_{bc}$ as independent variables with no symmetry requirements; that is, we take
$(\partial \Gamma^a_{bc}/\partial \Gamma^i_{jk})=\delta^a_i\delta^b_j\delta^c_k$ 
Obviously, the order of lower indices in $\Gamma^a_{bc}$ in \eq{sep} is important in arriving at this result.   
After the derivative is computed we will impose the condition that the $\Gamma^k_{bc}$ are related to the metric by the standard relation.
\bibitem{ps2}The $Q^{abcd}$ for the Einstein-Hilbert case is closely related to the superpotential for energy momentum pseudotensor for gravity. In fact, in the locally inertial frame, one can re-express
$L=Q^{abcd}R_{abcd}\sim Q^{abcd}\partial_a\partial_c g_{bd}$ in terms of $L_{mod}\sim g_{bd}\partial_a\partial_cQ^{abcd}\sim g_{bd}t^{bd} $
where $t^{bd} $ is the gravitational energy momentum pseudotensor.
\end{thebibliography}
\end{document}